\RequirePackage[2020-02-02]{latexrelease}
\documentclass[pre,aps,twocolumn,showpacs]{revtex4}
\usepackage{epsfig}
\include{graphics}

\begin{document}

\title{Spatial spread of infection: transitions between pulled and pushed fronts}

\author{Evgeniy Khain and Rohan Sukumar}
\email{khain@oakland.edu}
\affiliation{Department of Physics, Oakland University, Rochester, MI 48309, USA}

\begin{abstract}
We consider the spatial spread of epidemic into an unstable, healthy state. When the transmission rate depends on the fraction of infected, the propagating pulse of infection can be either a pulled front or a pushed front. We determined the phase space of parameters for the pulled and pushed regions both by numerically solving the spatial SIR partial differential equations and by a theoretical analysis of the front propagation phenomenon. We found both a continuous and a discontinuous transition between the pulled and pushed front solutions; in the latter transition, the front speed undergoes a jump as a certain parameter crosses the critical threshold. The behavior of the propagating pulses near the transitions has been analyzed and a good agreement between the theory and numerics has been observed. Finally, a bistable region where both pushed and pulled propagating pulses can be realized has been discovered.
\end{abstract}

\maketitle

\section{Introduction}

The phenomenon of front propagation is ubiquitous in nature, from the spread of invasive species \cite{invasive} and cell migration \cite{migration} to flame fronts propagating in reactive systems \cite{flame}, spread of wildfires \cite{fire} and autocatalytic chemical waves \cite{chemwaves}. Significant research has been done investigating fronts propagating into an unstable state \cite{Saarloos}. Such fronts are divided into pulled fronts, for which the front speed is determined by the leading edge of the front, and pushed fronts that generally move faster and whose dynamics are determined by the entire nonlinear front region \cite{Saarloos}. A continuous transition between pulled and pushed fronts has been observed when a governing parameter (let us denote it by $\mu$) crosses a certain threshold $\mu = \mu_c$ \cite{Saarloos,Korolev}, and near the transition, the front speed difference $c_{pushed}-c_{pulled}$ has been shown to scale quadratically with the distance to this threshold value $|\mu-\mu_c|$ \cite{Scheel}.

Analysis of propagating pulses of infection in the framework of the basic Susceptible-Infected-Recovered (SIR) model is a famous textbook problem \cite{book}. The linearized equation for the fraction of infected is analogous to the linearized Fisher-Kolmogorov equation \cite{FK1,FK2}, and the corresponding fronts move with the speed of $c_0=2D\sqrt{r-\alpha}$, where $D$ is the diffusion coefficient, $r$ is the transmission rate and $\alpha$ is the recovery rate. When $r>\alpha$, the state of no infection is linearly unstable, so the propagating pulse of infection is a pulled front moving into an unstable state. Are there pushed fronts of infection moving into an unstable state?

There are indications that the transmission rate may not be a constant, but a function of a fraction of infected \cite{Allee}. We have recently considered the SIR model with a modified (nonlinear) transmission rate and analyzed fronts propagating into a linearly stable state of no infection \cite{KhainPRE2023}. In this work, we adapt the same modified transmission rate and investigate pulses of infection propagating into an unstable, healthy state. We find both pushed and pulled fronts in different regions of the phase diagram of parameters and investigate the transitions between different front types. In addition to a continuous transition, we found discontinuous transitions and a region of bistability.

\section{The model}

The spatial susceptible-infected-recovered model for the fraction of susceptible $S(x,t)$, infected $I(x,t)$, and recovered individuals $R(x,t)$ \cite{book} is given by:

\begin{eqnarray}
\frac{\partial S}{\partial t} &=& - r S I +  D \frac{\partial^2 S}{\partial x^2}, \\ \nonumber
\frac{\partial I}{\partial t} &=& r S I - \alpha I  +  D \frac{\partial^2 I}{\partial x^2}, \\ \nonumber
\frac{\partial R}{\partial t} &=& \alpha I  +  D \frac{\partial^2 R}{\partial x^2}.
\end{eqnarray}

In the standard setting, the transmission rate $r$ is taken to be constant. However, there are indications that public health measures result in a lower transmission rate for a smaller fraction of infected and a higher transmission rate for a larger fraction of infected \cite{Allee}. Assuming that $r$ varies between $r = r_{min}$ for low $I$ to $r = r_{max}>r_{min}$ for high $I$, this dependence can be modeled as \cite{KhainPRE2023}
$$ r(I) = r_{min} + (r_{max}-r_{min})\frac{I}{\bar{I}+I}. $$ Note that Ref. \cite{KhainPRE2023} considered the case $r_{min} < \alpha$, where the state of no epidemic $S=1$, $I=0$ was stable, and the problem of front propagation into a stable state has been examined. This research assumes $r_{min} > \alpha$ and investigates the spatial propagation of a pulse of infection into an unstable (healthy) state.

Let us now introduce the dimensionless coordinate $\bar{x} = \sqrt{\alpha/D}\, x$ and the dimensionless time $\bar{t} = \alpha t$. The equations for the fractions of infected and susceptible become

\begin{eqnarray}
\frac{\partial S}{\partial \bar{t}} &=& - \bar{r} S I +  \frac{\partial^2 S}{\partial \bar{x}^2}, \\ \nonumber
\frac{\partial I}{\partial \bar{t}} &=& \bar{r} S I - I  +  \frac{\partial^2 I}{\partial \bar{x}^2},
\end{eqnarray}
where
\begin{equation}
\bar{r} = \bar{r}_{min} + (\bar{r}_{max}-\bar{r}_{min})\frac{I}{\bar{I}+I}
\end{equation}
with $\bar{r}_{min} = r_{min}/\alpha$ and $\bar{r}_{max} = r_{max}/\alpha$. Throughout this work, we will fix the value of $\bar{r}_{max}$ and vary the remaining two parameters: $\bar{r}_{min}>1$ and $\bar{I}$. As we show below, the propagating pulse of infection can be either a pulled front or a pushed front depending on the values of these two parameters.

\section{Phase diagram of parameters: pulled and pushed regions}

The speed of pulled fronts can be computed theoretically, as it is determined by the precursor and can be found from the linearization near the infection-free $I=0$ state. Indeed, substituting the front ansatz $ I = I(\xi = \bar{x}-c\bar{t})$ and linearizing the resulting equation in the vicinity of $I=0$ ($S=1$) state, we get:
$$-c\, \frac{d I}{d \xi} = \bar{r}_{min} I -  I  +  \frac{d^2 I}{d \xi^2}.$$ One can see a mechanical analogy with the damped harmonic oscillator, where the front speed plays a role of the damping coefficient. For small damping (small front speed $c$) the decay is oscillatory, which means that $I(\xi)$ will decay to zero in an oscillatory manner, so the fraction of infected will inevitably become negative for certain values of $\xi$, which is not allowed. Therefore, there should be a minimal speed, corresponding to the critical damping in our mechanical analogy. As in many other pulled front systems, sharp enough initial conditions develop in this case into a pulse of infection moving with this critical speed: $c_{pulled} = 2\sqrt{\bar{r}_{min}-1}$.

The numerical solution of Eqs. (2-3) in MATLAB shows that after a short transient, the profiles of $S$ and $I$ develop into fronts moving with a constant speed $c$. It is known that pushed fronts generally move faster $c_{pushed}>c_{pulled}$. Therefore, for every set of parameters, one can compare the resulting front speed with the value of $c_{pulled}$ and decide if this is a pulled or a pushed front. The resulting “pulled” and “pushed” regions on the ($\bar{r}_{min}$, $\bar{I}$) phase plane are shown in Figure 1. The border between the regions is shown both by black circles computed from the numerical solution of Eqs. (2-3) and by blue squares computed by employing the “shooting” numerical procedure (see the next section). Note also that the border between the two regions continues to the $\bar{r}_{min}<1$ part of the diagram where pulled fronts do not exist. There, above the threshold ($\bar{I} > \bar{I}_c$), the initially propagating pulse of infection slows down and decays \cite{KhainPRE2023}. In the opposite limit, the pushed front region disappears as $\bar{r}_{min}$ tends to $\bar{r}_{max}$. Indeed, when $\bar{r}_{min}=\bar{r}_{max}$, the transmission rate is constant, $\bar{r(I)} = \bar{r}_{max}$, and only pulled fronts can propagate in the system.

\begin{figure}[htp!]
\begin{center}
\includegraphics[width=3.5 in]{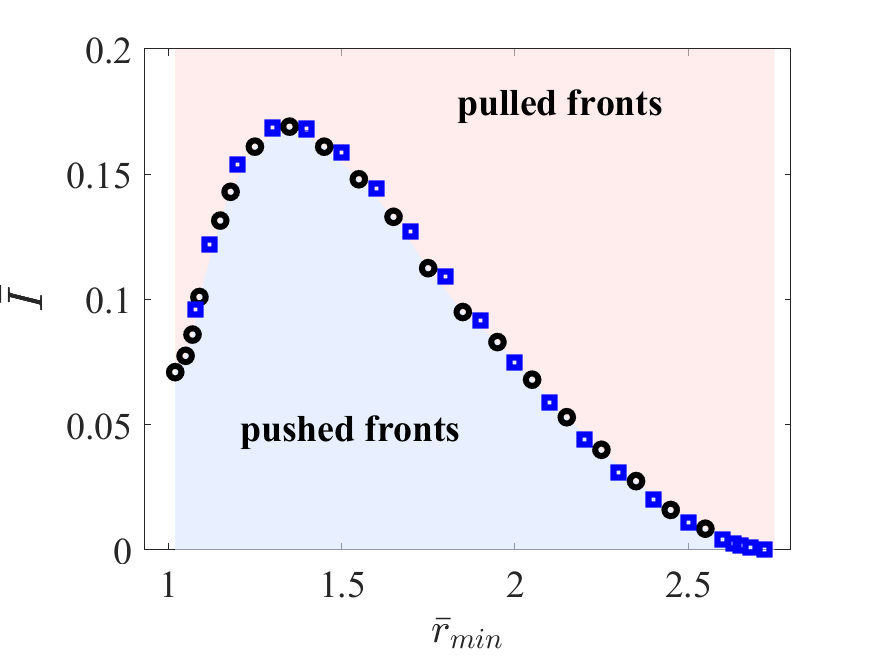}
\caption{Phase plane of parameters: the border between the regions of pulled and pushed fronts, $\bar{r}_{max} = 2.8$. Black circles are computed from the numerical solution of Eqs. (2-3). Blue squares are computed by employing the “shooting” numerical procedure, see text.
}
\end{center}
\end{figure}

Figure 2 shows an example of the pulled and pushed front profiles for both the fraction of infected (Fig. 2a) and the fraction of susceptible (Fig. 2b). As the transition across the border between the pulled and pushed regions can be discontinuous, the pulled and pushed front profiles can be remarkably different. In the pushed case (black dash-dotted lines), the amplitude of the infection pulse is higher (and therefore the remaining fraction of susceptible is lower) and the pulse decays faster compared to the pulled case (blue solid lines).

\begin{figure}[htp!]
\begin{center}
\includegraphics[width=3.5 in]{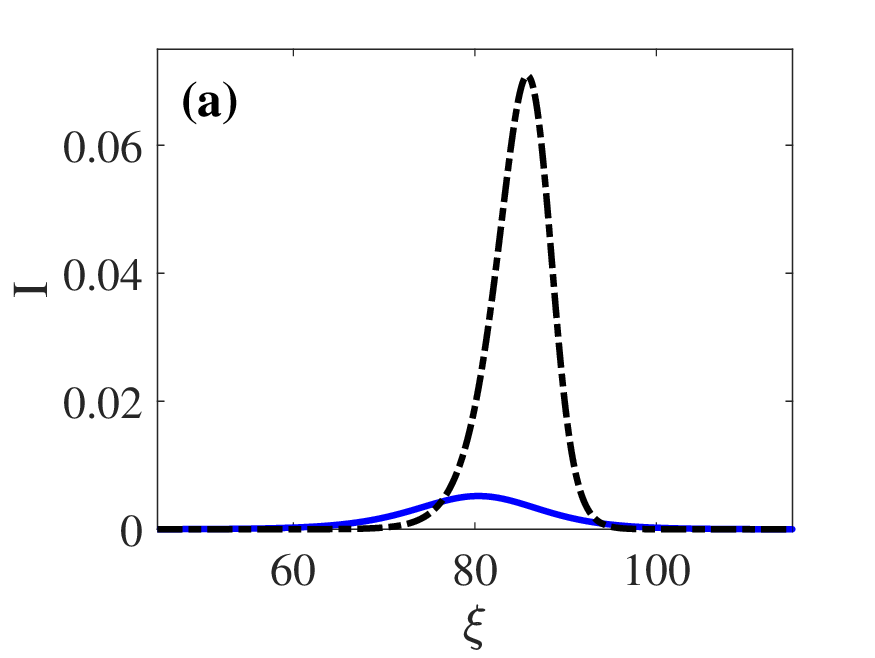}
\includegraphics[width=3.5 in]{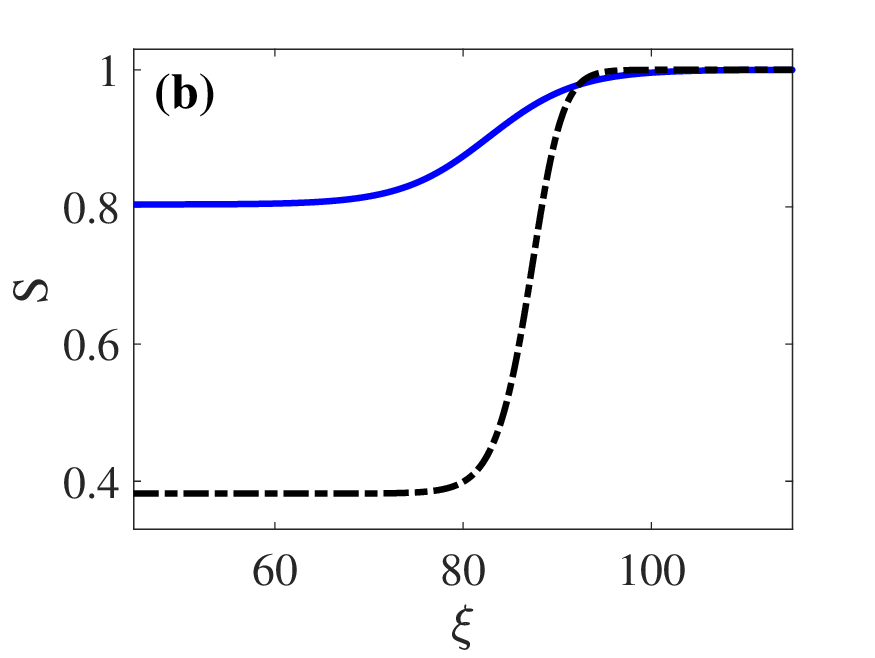}
\caption{The fraction of infected (a) and the fraction of susceptible (b): pulled and pushed front profiles. The parameters are $\bar{r}_{min} = 1.06$, $\bar{r}_{max} = 2.8$, $\bar{I}=0.0800$ (pulled front, blue solid line) and $\bar{I}=0.0745$ (pushed front, black dash-dotted line).
}
\end{center}
\end{figure}

\section{Theory of front propagation}

In order to better understand the behavior of the front speed near the border between the pulled and pushed regions, we employ a semi-theoretical approach. First, we substitute the front propagation ansatz $I = I(\xi)$ and $S = S(\xi)$ into Eqs. (2) and rewrite the two resulting equations as a four-dimensional dynamical system for $S$, $u=dS/d\xi$, $I$, and $v=dI/d\xi$. The front profile corresponds to the trajectory in this four-dimensional space connecting the state before the epidemic ($S=1$, $u=0$, $I=0$, $v=0$) with the state after the epidemic ($S=S_{final}$, $u=0$, $I=0$, $v=0$). Finding this trajectory is challenging as the values of both $S_{final}$ and $c$ are unknown a priori. To make progress, we analyze the system behavior near the two states and then employ a so-called “shooting” numerical procedure. Linearizing the system near the states ($S=S_*$, $u=0$, $I=0$, $v=0$), we obtain

\begin{eqnarray}
\frac{d (\delta S)}{d \xi} &=& u \\ \nonumber
\frac{d u}{d \xi} &=& -c\,u +   \bar{r}_{min} S_* I \\ \nonumber
\frac{d I}{d \xi} &=& v \\ \nonumber
\frac{d v}{d \xi} &=& -c\,v - \bar{r}_{min} S_* I +  I,
\end{eqnarray}
where $\delta S = S - S_*$, and $S_*=S_{final}$ (when considering the state left behind the front) or $S_*=1$ (when considering the state the front propagates to).

First, we analyze the behavior of $I$ and $S$ near the ($S=S_{final}$, $I=0$) fixed point. Demanding that $S$ approaches $S_{final}$ and $I$ approaches $0$ as $\xi$ tends to minus infinity, we find the approximate solution in the vicinity of the ($S=S_{final}$, $I=0$) state:
$$S(\xi) = S_{final} + \bar{d}\frac{\bar{r}_{min}\,S_{final}}{(c+\lambda_+)(\lambda_+)^2}\,\exp(\lambda_+ \xi)$$ and
$$I(\xi) = \frac{\bar{d}}{\lambda_+}\exp(\lambda_+ \xi),$$
where $\bar{d}$ is an arbitrary (small) constant and the relevant eigenvalue is
$$\lambda_+ = -c/2 + \sqrt{c^2/4 + 1 -\bar{r}_{min}S_{final}}.$$

Next, we study the behavior of $I$ and $S$ near the ($S=1$, $I=0$) fixed point. Again, there are four eigenvalues: $\lambda_1 = 0$, $\lambda_2 = -c$, $\lambda_3 = -c/2 + \sqrt{c^2/4 + 1 -\bar{r}_{min}}$, and $\lambda_4 = -c/2 - \sqrt{c^2/4 + 1 -\bar{r}_{min}}$. The general solution in the vicinity of $I=0$ is $I(\xi) = A_3\exp(\lambda_3\xi) + A_4\exp(\lambda_4\xi)$. This general solution, however, is realized neither for pushed fronts, nor for pulled fronts. Indeed, for pushed fronts, the steepest possible front is chosen and as $|\lambda_4| > |\lambda_3|$, $A_3=0$ (our numerical observations support this statement). As a result, $$I(\xi) = A_4\exp(\lambda_4\xi)$$ and $$S(\xi) = 1 + A_4\,\frac{\bar{r}_{min}}{(c+\lambda_4)\lambda_4}\,\exp(\lambda_4\xi) + A_c\exp(-c\xi).$$ On the other hand, for pulled fronts, $\lambda_3 = \lambda_4 \equiv \lambda_0$, so solution is written in the form $I(\xi) = A_1 \xi\exp(\lambda_0\xi) + A_0 \exp(\lambda_0\xi)$, where $A_1$ must be nonnegative to ensure that the fraction of infected individuals, $I$, remains nonnegative.

A standard “shooting” numerical procedure is employed to find $I(\xi)$ and $S(\xi)$ that satisfy the desired behavior in the vicinity of the two fixed points. In addition to the profiles of $I$ and $S$, the procedure provides values for the two a priori unknown parameters: the front speed $c$ and the fraction of susceptible $S_{final}$ behind the front. The phase diagram presented in Fig. 1 shows that for a fixed value of $\bar{r}_{min}$, a transition from pulled to pushed regions occurs as $\bar{I}$ is decreased. It turns out that this transition can be either continuous (for $\bar{r}_{min}$ above a certain critical value), where $c_{pushed} = c_{pulled}$ at the transition point, or discontinuous (for $\bar{r}_{min}$ below that critical value), where the front speed undergoes a jump.

\subsection{Continuous transition between pulled and pushed fronts}

Both the numerical solution of the time dependent equations (1) and the “shooting” numerical procedure employed to solve equations (4) show that for a fixed value of $\bar{r}_{min}$ pulled fronts exist for $\bar{I} > \bar{I}_c$ with the fraction of infected given by $I(\xi) = A_1 \xi\exp(\lambda\xi) + A_0 \exp(\lambda\xi)$. Below that threshold, the fronts are pushed and move with a larger speed, $c_{pushed} > c_{pulled}$. The coefficient $A_1$ is positive for the pulled region and becomes zero at the transition point \cite{Saarloos}. This is quite intuitive as $A_1=0$ corresponds to the steepest possible pulled front.

Figure 3 shows a continuous transition between the two front types: the speed of front propagation (blue circles) as a function of $\bar{I}$ for the fixed value of $\bar{r}_{min}$. One can observe a plateau for $\bar{I} > \bar{I}_c$ (the front speed is independent of $\bar{I}$ in the pulled region) and an increasing speed for $\bar{I} < \bar{I}_c$ (the pushed region). Near the transition (for small values of $\epsilon = \bar{I}_c - \bar{I}$), the front speed can be approximated by $c_{pushed} = c_{pulled} + \beta \epsilon^2$ \cite{Scheel}. This approximation is shown in Figure 2 by the blue dotted line, while the plateau for $\bar{I} > \bar{I}_c$ is shown by the red dashed line. Let us now derive this scaling and obtain the value of $\beta$.

\begin{figure}[htp!]
\begin{center}
\includegraphics[width=3.5 in]{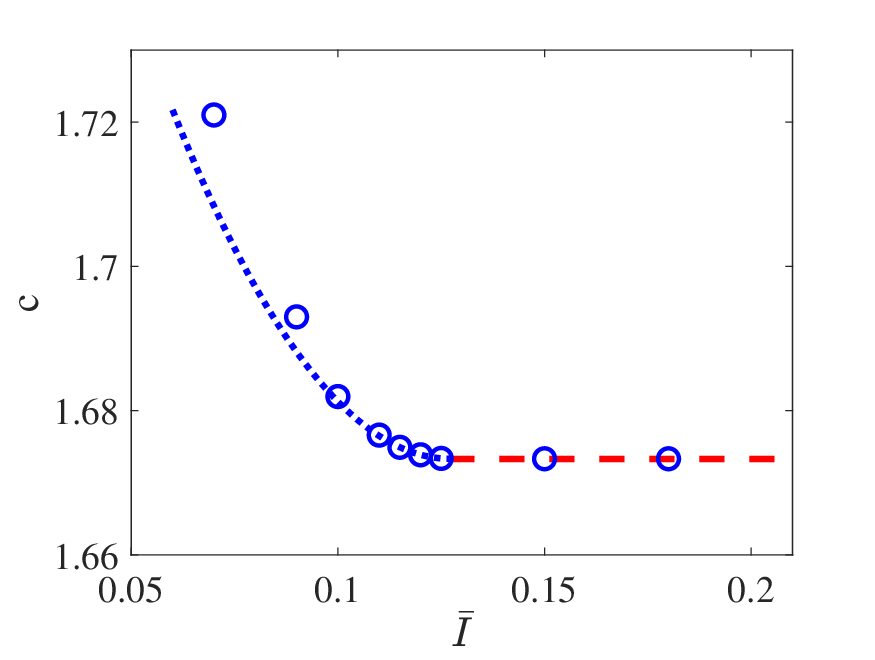}
\caption{Continuous transition: Front speed as a function of $\bar{I}$. The pushed front asymptotics is given by $c_{pushed} = c_{pulled} + \beta (\bar{I}_c - \bar{I})^2$, the blue dotted line, see text. The parameters are: $\bar{r}_{min} = 1.7$, $\bar{r}_{max} = 2.8$, $\beta \simeq 10.714$, $\bar{I}_c \simeq 0.127$.
}
\end{center}
\end{figure}

As $A_1=0$ for $\bar{I} = \bar{I}_c$, it generally should be proportional to $\epsilon$ near the transition, $A_1 = \alpha (\bar{I} - \bar{I}_c)$. We now employ a perturbation theory (valid at small $\epsilon$), assuming that the pushed front solution is approximately equal to the (non realized) pulled front solution with a small negative value of $A_1$:

$$A_4\exp((\lambda_0 - \delta)\xi) = A_1 \xi\exp(\lambda_0\xi) + A_0 \exp(\lambda_0\xi),$$ where
$$\delta \equiv \lambda_0 - \lambda_4 = \frac{1}{2} \left( c - c_{pulled} + \sqrt{c^2 - c^2_{pulled}}\right).$$
Assuming a small $\delta$ and expanding the exponent, we get $A_4 = A_0$ and $-A_4\delta = A_1$, so $\delta^2 = (c_{pulled}/2)(c-c_{pulled}) = (\alpha^2/A^2_0)\epsilon^2$ and $$c = c_{pulled} + \frac{2\alpha^2}{A^2_0 c_{pulled}}\epsilon^2.$$ The coefficient in front of $\epsilon$ can be obtained from the “shooting” numerical procedure for pulled fronts. Indeed, the ratio $v(\xi)/I(\xi)$ decreases with $\xi$ as $\lambda_0 + A_1/A_0 - (A_1/A_0)^2\xi$. We computed the slope $(A_1/A_0)^2$ above the transition, verified the scaling with $\epsilon$ and computed the coefficient $\beta = (2\alpha^2)/(A^2_0 c_{pulled})$. Figure 3 shows an excellent agreement of this theoretical scaling (the blue dotted line) with the observed front speed near the transition.

\subsection{Discontinuous transition between pulled and pushed fronts}

While Fig. 3 shows a continuous transition between the speeds of pulled and pushed fronts, for smaller values of $\bar{r}_{min}$, the transition is discontinuous, and the front speed undergoes a jump, see Fig. 4. Above the critical value of $\bar{I}$ ($\bar{I} > \bar{I}_c$), the pulled front is the only existing solution (the theoretical black dash-dotted line), while below the threshold ($\bar{I} < \bar{I}_c$), there are two additional pushed front branches, shown by the red circles (a stable branch) and the blue circles (an unstable branch) and computed employing the “shooting” numerical procedure. Figure 4 shows that the pushed front branches undergo a saddle-node bifurcation, which means that near the transition (for small values of $\epsilon = \bar{I}_c - \bar{I}$), the pushed front speed of the stable branch (the red solid line) is given by $c_{pushed} = c_{crit} + B \epsilon^{1/2}$ ($c_{pushed} = c_{crit} - B \epsilon^{1/2}$ for the unstable branch, the blue dotted line).

To theoretically derive this approximation, we focus on the pushed front solution. While the general solution is given by $I(\xi) = A_3\exp(\lambda_3\xi) + A_4\exp(\lambda_4\xi)$, for pushed fronts for any fixed $\bar{r}_{min}$, the coefficient $A_3(c,\bar{I})$ must be equal to zero. For a saddle-node bifurcation, near the transition, $$A_3 = -B_1(c-c_{crit})^2 + B_2(\bar{I}_c - \bar{I}).$$ This behavior is verified in Fig. 5. The “shooting” numerical procedure allows computing $A_3$ for any value of $c$ and $\bar{I}$ just by following the large $\xi$ limit of the product $I(\xi) \exp(-\lambda_3\xi)$. Figure 5 shows the coefficient $A_3$ as a function of $c$ for three values of $\bar{I}$: below the transition (black circles, two roots), (almost) at the transition (blue squares, a single root), and above the transition (magenta diamonds, no roots). A fit to all three curves gives $c_{crit}=0.729$, $\bar{I}_c=0.07491$, $B_1 = 0.0765$, and $B_2 = 0.5$. Demanding $A_3=0$ produces $c_{pushed}(\epsilon)$, shown in Fig. 4 by the red solid line for a stable branch and by the blue dotted line for an unstable branch.

\begin{figure}[htp!]
\begin{center}
\includegraphics[width=3.5 in]{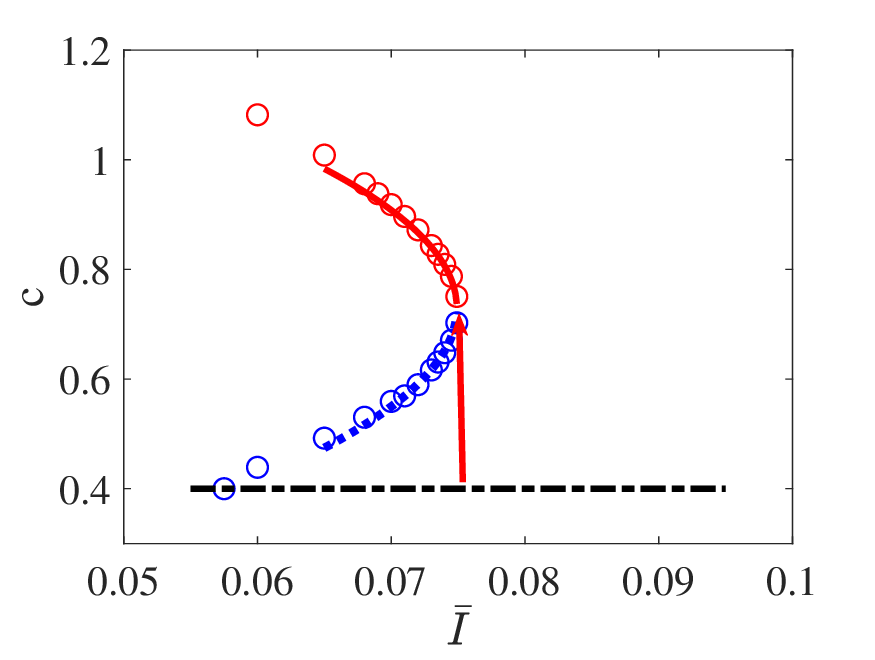}
\caption{The front speed as a function of $\bar{I}$ for $\bar{r}_{min}$ below the critical threshold: a discontinuous transition. The circles are computed by employing the “shooting” numerical procedure, while the red solid line and the blue dotted line represent the theoretical approximation near the bifurcation, see text. The black dash-dotted line describes the theoretical pulled front speed. $\bar{r}_{min} = 1.04$, $\bar{r}_{max} = 2.8$.
}
\end{center}
\end{figure}

\begin{figure}[htp!]
\begin{center}
\includegraphics[width=3.5 in]{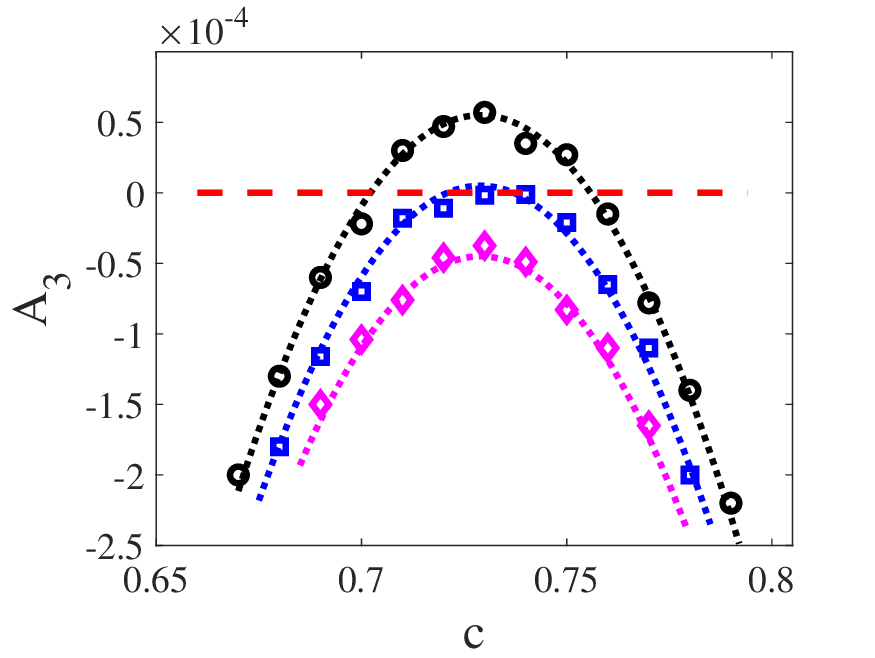}
\caption{$A_3$, the coefficient in front of the slowly decaying exponent, as a function of $c$ for three values of $\bar{I}$. The pushed front solution requires $A_3=0$, therefore, one can observe a saddle-node bifurcation as $\bar{I}$ is varied. $\bar{I} = 0.0748$ (black circles), $\bar{I} = 0.0749$ (blue squares), and $\bar{I} = 0.0750$ (magenta diamonds). $\bar{r}_{min} = 1.04$, $\bar{r}_{max} = 2.8$, see text for the theoretical fit (dotted lines).
}
\end{center}
\end{figure}

\begin{figure}[htp!]
\begin{center}
\includegraphics[width=3.5 in]{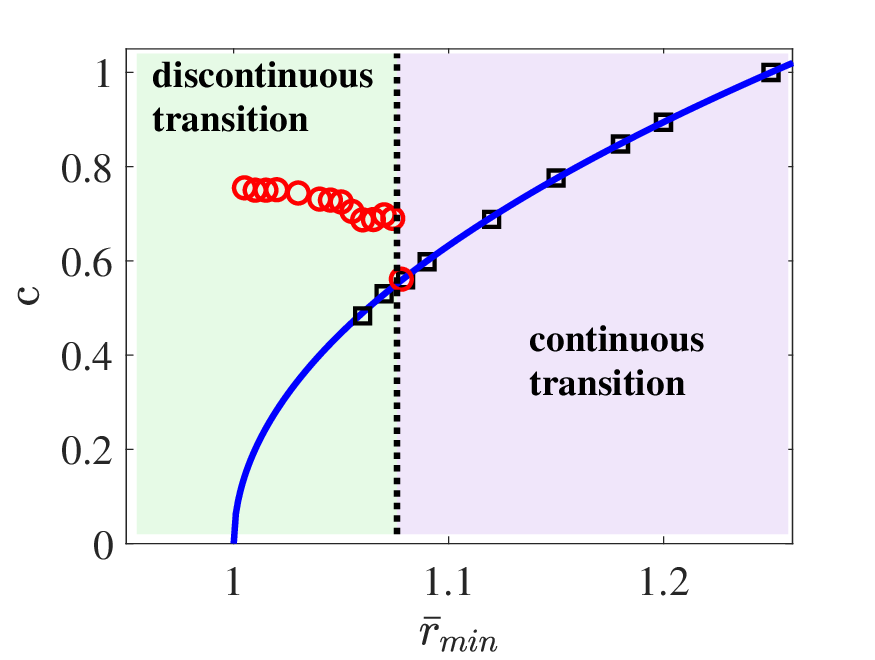}
\caption{The pushed and the pulled front speeds as a function of $\bar{r}_{min}$ along the border (and from both sides of the border) of the phase diagram (Fig. 1). Red circles denote the speeds of the pushed fronts, while black squares denote the speeds of the pulled fronts as computed from the numerical solution of Eqs. (1). The theoretical pulled front speed is shown by the blue solid line. The vertical dotted line at $\bar{r}_{c}=1.076$ separates the regions of discontinuous (left) and continuous (right) transitions. $\bar{r}_{max} = 2.8$.
}
\end{center}
\end{figure}

\section{Bistability}

Figure 6 illustrates both the continuous and discontinuous transitions showing the pushed and the pulled front speeds as a function of $\bar{r}_{min}$ along the border (and from both sides of the border) of the phase diagram (Fig. 1). The blue solid line corresponds to the theoretical pulled front speed, $c_{pulled} = 2\sqrt{\bar{r}_{min}-1}$. The symbols are computed from the numerical solution of Eqs. (1): red circles denote the pushed front speed, while black squares denote the speeds of the pulled fronts. The vertical dotted line $\bar{r}_{min} = \bar{r}_{c} = 1.076$ divides the diagram into two regions. The region on the right corresponds to the continuum transition, so both the pushed and pulled front speed along the border equal to $c_{pulled}$. The region on the left corresponds to a discontinuous transition, so the front speed jumps from $c_{pulled}$ to $c_{pushed}$ as $\bar{I}$ crosses the critical threshold for the fixed $\bar{r}_{min}$ as shown in Fig. 1.

Let us focus on the $\bar{r}_{min}<\bar{r}_{c}$ region in more detail. Figure 4 shows that the pushed front solution does not exist for $\bar{I}>\bar{I}_c$, but below this threshold both the pulled and pushed fronts do formally exist. The numerical solution of Eqs. (1) with initial conditions $I(x,t=0) = 0.06$ for $x<0$ and $I(x,t=0) = 0$ for $x>0$ shows that the transient dynamics always lead to a pushed front. However the basin of attraction of the pulled front solution can be small but nonzero. To test this hypothesis, we started the numerical simulations in the pulled region for $I>\bar{I}_c$, followed the transient dynamics that leads to a pulled front and then switched to $\bar{I}<\bar{I}_c$. Quite remarkably, we did find the pulled front solution when the switch $\Delta\bar{I}$ was not too big. It means that one can realize both the pushed front solution and the pulled front solution for the same set of parameters! The region of bistability in the ($\bar{r}_{min}$ $\bar{I}$) phase diagram is too small to be shown in Fig. 1. For example, for $\bar{r}_{min}=1.076$ in the discontinuous region (see Figs. 1 and 6), the pushed front solution does not exist for $\bar{I}>0.080$, so there are only pulled fronts in this region of parameters. Only pushed front solutions are realized for $\bar{I}<0.075$, but one can obtain both propagating pulled and pushed pulses of infection in the interval $0.075<\bar{I}<0.080$. These two fronts are entirely different: they move with different speeds and the amplitude of the pulse of infection in the pushed case is much larger compared to that in the pulled case.

\section{Summary and Discussion}

We investigated the propagation of a pulse in infection into an unstable, healthy state in the framework of the SIR epidemiological model with a nonlinear transmission rate. Initial outbreak develops to a propagating front, which can be either pulled or pushed depending on the regime of parameters. We presented the entire phase diagram of parameters identifying the pulled and pushed regions and studied the transitions between the pulled and pushed front solutions. We observed both the continuous transition, where $c_{pushed} = c_{pulled}$ at the critical value of a parameter, and a discontinuous transition, where the front speed undergoes a jump, so $c_{pushed} > c_{pulled}$ at the transition. In addition, we discovered a region of bistability, where depending on initial conditions, both the pushed front and the pulled front can be realized for the same governing parameters. Our numerical results computed by solving the system of partial differential equations fully agree with our theoretical results obtained from the analysis of the front propagation problem.

The phenomena discussed in this work result from the dependence of the transmission rate on the fraction of infected. If $\bar{I}=0$, the transmission rate is constant and equals its maximum value, $r = r_{max}$. In the model, public health measures lead to a nonzero $\bar{I}$, resulting in a smaller transmission rate in the beginning of the epidemics. Although the epidemic outbreak (the pulse of infected) will propagate through the system for any $\bar{I}$, the phase diagram in Fig. 1 shows that larger $\bar{I}$ correspond to pulled fronts that propagate slower and infect a smaller fraction of the population. The difference can be quite striking in the discontinuous transition region, see Fig. 2. Although this is clearly a toy model and we do not know how exactly the transmission rate depends on the fraction of infected, the qualitative result is quite encouraging: public health measures (which increase $\bar{I}$) can be quite helpful.

An interesting avenue of future research is investigating the role of stochastic fluctuations in this system due to intrinsic shot noise that is known to affect front propagation \cite{MSK}. These fluctuations can be very important in the bistability region of the phase diagram and generally produce corrections to the front speed \cite{fluctuations}.

\end{document}